%Paper: hep-lat/9512013
%From: S.Hands@swansea.ac.uk
%Date: Fri, 08 Dec 95 11:11:46 +0000
%Date (revised): Fri, 08 Dec 95 12:12:31 +0000

%Paper: hep-lat/9512013
%From: S.Hands@swansea.ac.uk
%Date: Fri, 08 Dec 95 11:11:46 +0000

%%%%%%%%%%%%%%%% tex file follows %%%%%%%%%%%%%%%%%%%
\magnification = 1200
\baselineskip=16pt
\vbadness=5000
\null\vskip-30pt
\rightline{SWAT/95/93}
\rightline{hep-lat/yymmxxx}
\rightline{December 1995}
\vskip0.5 truecm
\centerline{\bf MONTE CARLO SIMULATION OF THE THREE DIMENSIONAL THIRRING
MODEL}
\vskip 0.5 truecm
\centerline{{\bf Luigi Del Debbio and Simon Hands}}
\centerline{\it Department of Physics}
\centerline{\it University of Wales, Swansea,}
\centerline{\it Singleton Park,}
\centerline{\it Swansea SA2 8PP, U.K.}
\vskip 1.5 truecm
\centerline{\bf Abstract}

{\narrower
\noindent
We study the Thirring model in three spacetime dimensions, by means of
Monte Carlo simulation on lattice sizes $8^3$ and $12^3$,
for numbers of fermion flavors $N_f=2,4,6$. For
sufficiently strong interaction strength, we find that spontaneous
chiral symmetry breaking occurs for $N_f=2,4$, in accordance with
the predictions of the Schwinger-Dyson approach. The phase transitions
which occur are continuous and with critical scaling behaviour depending
on $N_f$.
For $N_f=6$ our results are preliminary, and no firm conclusions about
the existence or otherwise of chiral symmetry breaking are possible.


\smallskip}
\vfill\eject
\baselineskip=20pt
\parskip= 5pt plus 1pt
\parindent=15pt

The three dimensional Thirring model is a field theory of relativistic
fermions interacting via a contact term between conserved vector
currents. Its Lagrangian is written
$${\cal L}=\bar\psi_i(\partial{\!\!\! /}\,+m)\psi_i
+{g^2\over{2N_f}}(\bar\psi_i\gamma_\mu\psi_i)^2,\eqno(1)$$
where $\psi_i,\bar\psi_i$ are four-component spinors, $m$ is a bare,
parity-conserving mass, and the index $i$ runs over $N_f$ distinct
fermion species. Since the coupling $g^2$ has mass dimension $-1$, naive
power-counting suggests that the model (1) is non-renormalisable.
However, as has been suspected for many years [1,2], an
expansion in powers of $1/N_f$, rather than $g^2$, is exactly
renormalisable. At leading order in $1/N_f$, interaction between vector
currents is dominated by exchange of a fermion - anti-fermion bound
state described in terms of a chain of vacuum-polarisation ``bubbles''.
In the ultra-violet limit the interaction is thus transformed from a
monmentum-independent contact term
to a softer $A/(N_f k)$ behaviour, where $A$ is a
numerical constant independent of $g$: this asymptotic behaviour can be
used to evaluate divergent graphs at higher orders in $1/N_f$, eg. in
[3], where renormalisability of the massless model is explicitly
demonstrated to $O(1/N_f)$.

The property of renormalisability signals that the model's $1/N_f$
expansion exhibits a UV-stable fixed point of the renormalisation group,
the continuum limit being taken in the limit $g^2\Lambda\to\infty$,
where $\Lambda$ is a UV cutoff. RG fixed points have also been observed
in other three-dimensional four-fermi models [4,5]. The distinctive
feature of the Thirring model is that for $d<4$ the vacuum polarisation
is UV-finite so long as the regularisation respects current conservation
(eg. Pauli-Villars). This means that there is
no need to fine-tune $g^2$ to a critical value: a continuum limit may be
taken for any value of the dimensionless parameter $mg^2$ (at least to
leading order in $1/N_f$ [3]), the theory thus obtained having a
variable ratio of, say, physical fermion mass to vector bound state mass.
In the RG sense the interaction $(\bar\psi\gamma_\mu\psi)^2$ is a
{\it marginal\/} operator for $2<d<4$, whereas, say, the interaction
$(\bar\psi\psi)^2$ in the Gross-Neveu model is {\it relevant\/} [6].

Another possibility raised by the $1/N_f$ expansion is the equivalence
of the Thirring model in the strong-coupling limit $mg^2\to\infty$,
in which the vector particle becomes massless, with
the infra-red limiting behaviour of QED in three spacetime dimensions.
In massless large-$N_f$ ${\rm QED}_3$ vacuum polarisation screens
one-photon exchange to the extent that the $1/k^2$ interaction is again
transformed to $1/k$ [7]. The $O(1/N_f)$ corrections to the models
evaluated in respectively UV (Thirring) or IR (QED) limits appear to
coincide [3,8].

The $1/N_f$ expansion may not, however, describe the true behaviour of
the model, particularly for small $N_f$. For instance, spontaneous
chiral symmetry breaking, signalled by a vacuum condensate
$\langle\bar\psi\psi\rangle$, is forbidden to all orders in $1/N_f$, and
yet may be predicted by a self-consistent approach such as solution of
the Schwinger-Dyson equations [2,9,10]. In this approach a
non-trivial solution for the dressed fermion propagator
$S(p)=(A(p)ip{\!\!\! /}\,+\Sigma(p))^{-1}$ is sought, ie. one in which
the self-energy $\Sigma(p)$ and hence $\langle\bar\psi\psi\rangle$ are
non-vanishing in the chiral limit $m\to0$. Unfortunately, the SD
equations can only be solved by truncating them in a somewhat arbitrary
fashion. The usual approximation [2,10] is to assume that the
vector propagator is given by its leading-order form for $m=0$ in the
$1/N_f$ expansion, viz.
$$D_{\mu\nu}(k)=\left(\delta_{\mu\nu}-{{k_\mu k_\nu}\over k^2}\right)
\left(1+{{g^2(k^2)^{1\over2}}\over8}\right)^{-1}
+{{k_\mu k_\nu}\over k^2},\eqno(2)$$
and that the fermion-vector vertex function is well-approximated by the
bare vertex (the so-called ``planar'' or ``ladder'' approximation):
$$\Gamma_\mu(p,q)=-{{ig}\over\sqrt{N_f}}\gamma_\mu.\eqno(3)$$

The longitudinal part of $D_{\mu\nu}$ raises potential ambiguities: the
most systematic treatment has been given by Itoh {\it et al\/} [10],
who note the equivalence of the Thirring model with a gauge-fixed form
of a fermion-scalar model possessing a local gauge symmetry and then use
a non-local gauge-fixing condition to find a gauge in which the
``wavefunction renormalisation'' $A(p)\equiv1$. This simplification
enables the SD equations to be exactly solved in the limit
$g^2\to\infty$, with the result that a non-trivial solution for
$\Sigma(p)$ exists for
$$N_f<N_{fc}={128\over{3\pi^2}}\simeq4.32.\eqno(4)$$
Moreover since the integral equations require the introduction of a UV
cutoff $\Lambda$, a feature of this solution is that the induced physical mass
scale $\mu$ depends on $N_f$ in an essentially singular way:
$${\mu\over\Lambda}\propto\exp\left(-{2\pi\over\sqrt{{N_{fc}\over
N_f}-1}}\right);\;\;\;
\langle\bar\psi\psi\rangle\propto\Lambda^{1\over2}\mu^{3\over2}\propto
\exp\left(-{3\pi\over\sqrt{{N_{fc}\over
N_f}-1}}\right).\eqno(5)$$
This implies that a continuum limit only exists as $N_f\to N_{fc}$, the
scenario being very similar to that proposed by Miranskii and
collaborators for strongly-coupled ${\rm QED}_4$ [11]. Unfortunately no
analytic solution exists for $g^2<\infty$; however using different
techniques Kondo [12] has argued that a critical line $N_{fc}(g^2)$
exists in the $(g^2,N_f)$ plane, which is a smooth invertible function.
Therefore for integer $N_f<N_{fc}$ one might expect a critical scaling
behaviour
$$\langle\bar\psi\psi\rangle\propto\exp\left(-{a\over\sqrt{{g^2\over
g_c^2}-1}}\right),\eqno(6)$$
corresponding to a symmetry restoring transition at some critical point
$g^2=g_c^2$. Presumably in this scenario the Thirring interaction has
become relevant: there may exist a novel strongly-coupled continuum
limit at the critical point not described by the $1/N_f$ expansion.

There are good reasons to be cautious of this picture, however. Using a
different sequence of truncations Hong and Park have found chiral
symmetry  breaking for all $N_f$ [9], with
$${1\over g_c^2}\propto\exp\left(-{{N_f\pi^2}\over16}\right),\eqno(7)$$
a result which is non-analytic in $1/N_f$. Moreover in the limit
$g^2\to\infty$ the system of SD equations obtained are very similar to
those of large-$N_f$ ${\rm QED}_3$, in which case studies beyond the
planar approximation, using improved ans\"atze for the vertex
$\Gamma_\mu$, suggest that the condition $A(p)\equiv1$ is unphysical,
and that chiral symmetry is spontaneously broken for all $N_f$ [13].
In the current context this would imply $g_c^2<\infty$ for all $N_f$.

For these reasons we consider a numerical study of the
lattice-regularised model to be timely. If, as suggested above, the
Thirring model lies in the same universality class as ${\rm QED}_3$,
then a numerical study may shed light on the value of $N_{fc}$ for this
model [7]; previous lattice studies [14] have been plagued by large
finite volume effects due to the slow fall-off of the photon propagator
$\propto1/x$. The corresponding propagator in the Thirring model falls
as $1/x^2$, so sensible results may emerge on smaller systems. A second
motivation is the possible existence of a novel continuum limit: since
vacuum polarisation corrections to the vector propagator are finite to
all orders in $1/N_f$ there should be no competing effects of charge
screening, which obscures the issue in ${\rm QED}_4$.

The lattice action we have used is as follows:
$$\eqalign{
S&={1\over2}\sum_{x\mu i}\bar\chi_i(x)\eta_\mu(x)
\left(\chi_i(x+\hat\mu)-\chi_i(x-\hat\mu)\right)
+m\sum_{xi}\bar\chi_i(x)\chi_i(x)\cr
&\phantom{
={1\over2}
}
+{g^2\over{2N}}\sum_{x\mu ij}\bar\chi_i(x)\chi_i(x+\hat\mu)
\bar\chi_j(x+\hat\mu)\chi_j(x)\cr
&={1\over2}\sum_{x\mu i}\bar\chi_i(x)\eta_\mu(x)\left(1+iA_\mu(x)\right)
\chi_i(x+\hat\mu)+{\rm h.c.}+m\sum_{xi}\bar\chi_i(x)\chi_i(x)
+{N\over{4g^2}}\sum_{x\mu}A_\mu^2(x),\cr
&={1\over2}\sum_{xy\mu i}\bar\chi_i(x)M_{(A,m)}(x,y)
\chi_i(y)
+{N\over{4g^2}}\sum_{x\mu}A_\mu^2(x),\cr
}\eqno(8)$$
where $\chi,\bar\chi$ are staggered fermion fields, $\eta_\mu$ the
Kawamoto-Smit phases, $m$ is the bare mass, the flavor index $i$ runs
from 1 to $N$, and we have introduced $M_{(A,m)}$ for the fermionic
bilinear, which
depends on both the auxiliary field and the mass. The
second form of the action is the one actually simulated: the equivalence
of the two forms follows from Gaussian integration over the real-valued
auxiliary field $A_\mu$ defined on the lattice links (for $N=1$ there is
an alternative formulation in terms of a compact complex-valued
auxiliary [15,16]). The vector-like interaction of the action allows the
introduction of a checker-board, which in turn enables simulation of
the system for any $N$.
In three Euclidean dimensions staggered
fermions describe two continuum species of four-component fermions, with
a parity-conserving mass term [17]. Hence the number of physical flavors
$N_f=2N$. An interesting feature of the lattice formulation (8) is that
the interaction current is not exactly conserved. The conserved current
in lattice gauge theory incorporates the gauge connection
$\exp(iA_\mu)$. This means that at leading order in $1/N$ the vector
propagator receives an extra contribution from vacuum polarisation,
essentially due to the absence of the diagram of Fig. 1. The effect
can be absorbed into a redefinition of the coupling:
$$g_R^2={g^2\over{1-g^2J(m)}},\eqno(9)$$
where $J(m)$ is the value of the integral depicted in Fig. 1. The
physics described by continuum $1/N_f$ perturbation theory occurs for
the range of couplings $g_R^2\in[0,\infty)$, ie. for $g^2\in[0,g_{lim}^2)$;
to leading order in $1/N$
$${1\over g_{lim}^2}=J(m);\;\;\;{\rm
with}\;\;\;J(0)={2\over3}.\eqno(10)$$
We therefore expect to see some kind of discontinuous behaviour in our
simulations for small values of $1/g^2$.

In this letter, we aim to clarify the chiral symmetry breaking pattern
by studying the chiral condensate:
$$
\langle\bar\psi\psi\rangle = {1\over V}{\rm Tr } \left(M_{(A,m)}^{-1}\right)
$$
which, in the limit $m\to0$, is an order parameter for the spontaneous
symmetry breaking.
We performed simulations on $8^3$ and $12^3$ lattices for $N_f=2,4,6$,
using the hybrid Monte Carlo algorithm. Bare mass values $m$ ranged from
0.4 down to 0.02, with most attention paid to the range 0.05 -- 0.02.
For each mass and coupling we performed roughly 500 HMC trajectories,
the trajectory length being drawn from a Poisson distribution with mean
0.9. The condensate $\langle\bar\psi\psi\rangle$ was measured with
a stochastic estimator every few trajectories.
To maintain reasonable acceptance rates the timestep varied from
0.15 on $8^3$ at $m=0.4$ down to 0.022 on $12^3$ at $m=0.02$. We found
that considerably more work was needed to perform matrix inversion in
this model than for the Gross-Neveu simulations described in [5].
Another difference is that in this case since the critical region of
interest occurs at successively stronger couplings as $N_f$ is raised,
the {\it CPU\/} required also grows with $N_f$, despite the $1/N_f$ suppression
of quantum corrections.

In Fig.~2, we plot $\sigma\equiv\langle\bar\psi\psi\rangle$ vs. $1/g^2$ for the
three values of $N_f$ studied, for $m=0.10$ on a $8^3$ lattice. The models
with different $N_f$ have apparently coincident condensates in the
strong--coupling region $1/g^2\le 0.3$, but thereafter the
$\langle\bar\psi\psi\rangle$ signals peak to maxima at distinct values of
$1/g^2$ before falling away. It is tempting to associate the
strong--coupling region with $g_R^2<0$ from the discussion following
Eq.~(9), although the correspondence with the value of $g_{lim}^2$ predicted
in Eq.~(10) is not good. It may well be that the value of the diagram of
Fig.~1 is considerably altered in a chirally broken vacuum.

As stated above, in order to study spontaneous chiral symmetry breaking, one
has
to monitor the value of $\langle\bar\psi\psi\rangle$ as $m\to0$.
Our results for the chiral condensate for $N_f=2$ are reported in Fig.~3
for different values of the bare mass.
A naive
extrapolation to the chiral limit from the lattice data at fixed $1/g^2$
is probably  unreliable in the range of bare masses we have explored.
In order to determine the critical point, we need to perform a global fit of
our data incorporating many values of $m$ and $1/g^2$.
Therefore, we have to use an
equation of state (EOS) relating the external
symmetry breaking parameter $m$ to the response of the system
$\langle\bar\psi\psi\rangle$ and the coupling $1/g^2$ [16,18].

A generic EOS, inspired by the critical behaviour of spin systems,
can be written in terms of the scaled variables:
$$
m \langle\bar\psi\psi\rangle^{-\delta} = {\cal F}\left(\Delta(1/g^2)
\langle\bar\psi\psi\rangle^{1/\beta}\right), \eqno(11)
$$
where $\Delta(1/g^2)=1/g_c^2-1/g^2$ is the reduced coupling.
At $g=g_c$, Eq.~(11) is the usual scaling relation:
$$
\langle\bar\psi\psi\rangle \sim m^{1/\delta},
$$
while a Taylor expansion for small $\Delta(1/g^2)$ yields:
$$
m = B \langle\bar\psi\psi\rangle^{\delta} + A
\Delta(1/g^2) \langle\bar\psi\psi\rangle^{\delta-1/\beta} +\ldots
\eqno(12)
$$
where $A,B={\cal F}^{\prime}(0),{\cal F}(0)$ respectively.
At this stage, one sees that, for vanishing $m$, Eq.~(12) is simply
the definition of the critical exponent $\beta$:
$$
\langle\bar\psi\psi\rangle \sim (1/g_c^2-1/g^2)^{\beta},
$$
and that there are no logarithmic corrections, since these should only
appear in 4-d~[18].
If the critical behaviour is described by mean-field theory, we expect
$\delta=3$ and $\beta=1/2$, yielding:
$$
\langle\bar\psi\psi\rangle^2 = \kappa_1 {m \over \langle\bar\psi\psi\rangle}
+ \kappa_2 \Delta(1/g^2),
\eqno(13)
$$
which shows that $\langle\bar\psi\psi\rangle^2$ is a linear function of the
ratio $m/\langle\bar\psi\psi\rangle$. Such a plot is known as
Fisher plot. From Eq.~(13)  we see that a positive value of the intercept
corresponds to a non-vanishing value of the chiral condensate for $m=0$,
while the intercept will be exactly zero at the critical coupling.
In Fig.~4 and 5 we show the Fisher plots for $N_f=2,4$ respectively, where
we can see at first glance an indication of chiral symmetry breaking,
according to the criterion stated above.
In order to get a more quantitative evidence, we have fitted our
data using Eq.~(12) and a simpler version of it based on the hypothesis
that $\delta-1/\beta=1$ [19], which we will call respectively fit I and II in
what follows. We should stress that in the absence of a systematic
critical theory the forms I and II are used simply as effective
descriptions of the data.
The values of the fit parameters, their errors and the $\chi^2$ are listed
in Table 1. The number of values of the chiral condensate
included in the fit
is chosen in order to minimize the value of the reduced $\chi^2$. The
results of fit II are shown in Figs.~4 and 5 and seem to describe the
data quite well. The dashed line in Fig.~3 is the curve one obtains using
the results of fit II with
$m=0$. It shows clearly that, within the range
of $m$ we have explored, the chiral condensate is still far from its chiral
limit value, thus providing a justification {\it a posteriori\/} for the
impossibility of extrapolating the data naively.

There are a few conclusions one can draw from the numerical analysis that we
would like to emphasize.
First, for both values of $N_f$, we find clear evidence of chiral symmetry
breaking
at finite values of the coupling, as predicted by the Schwinger-Dyson
approach. From Eq.~(7) we get:
$$
{g^2_c(N_f=2) \over g^2_c(N_f=4)} = \exp(-\pi^2/8) = 0.291
$$
which is not too far from the fitted value $0.342 \pm 0.015$ (using the
data from fit II).

Secondly, although any claims to understand the details of the critical
scaling must be premature, the fits strongly suggest that the models with
$N_f=2$ and $4$ are described by distinct critical theories, in a sense
on ``either side'' of the mean field theory. If we relax the requirement
$\delta-1/\beta=1$ (which means using fit I instead of fit II)  then the
difference in the fitted values of
$\delta$ becomes even more apparent. This is significant because similar
EOS fits in ${\rm QED}_4$ reveal no such differences between $N_f=2$ and
$N_f=4$~[20].

Finally, we report some preliminary results for $N_f=6$. Figure~6 shows
$\langle\bar\psi\psi\rangle$ vs. $m$ for two values of $g$.
The $1/g^2=0.5$ data suggest a linear extrapolation to the chiral limit,
yielding a small condensate equal to 0.013(4). For $1/g^2=0.4$ it
is less clear how to make the extrapolation. Clearly in either case
reliable data
at much smaller mass values would be needed for confirmation or
otherwise of chiral symmetry breaking:
comparing data from different lattice sizes, we have found
that finite size effects become more important as we go to larger
$N_f$, which means that larger lattice sizes will probably be needed
before we can proceed to a more
quantitative study.

We conclude by briefly summarizing our results. We have shown by
numerical simulations that spontaneous chiral symmetry breaking does occur
in the Thirring model for finite $N_f$, in contradiction with the $1/N_f$
perturbative expansion, but in agreement with the Schwinger-Dyson approach
at least
for $N_f=2,4$. In these two cases we were able to determine the critical
coupling and the critical exponent $\delta$ by fitting to a plausible EOS,
and the fits suggest that the two models have different critical
behaviour. For
$N_f=6$, we were unable to find clear evidence in favour of symmetry
breaking, but cannot yet exclude a non-vanishing condensate in the chiral
limit.
In the future we plan to investigate in more detail the
theory for $N_f=2,4,6$ at the critical point,  focussing on critical
exponents, the renormalized charge, and spectroscopy.

{\bf Acknowledgments}

\noindent
LDD is supported by an EC HMC Institutional Fellowship under contract No.
ERBCHBGCT930470, and SJH by a PPARC Advanced Fellowship. Some of the
numerical work was performed on the Cray Y-MP at Rutherford--Appleton
Laboratory under PPARC grant GR/J675.5. We have enjoyed discussing this
project with Ji\v r\'\i ~Jers\'ak, Kei-Ichi Kondo,
Mike Pennington, Craig Roberts and Paolo Rossi.

\vfill\eject
\null
\noindent{\bf References}

\noindent
[1] G. Parisi, Nucl. Phys. {\bf B100} (1975) 368;\hfill\break
S. Hikami and T. Muta, Prog. Theor. Phys. {\bf57} (1977)
785;\hfill\break
Z. Yang, Texas preprint UTTG-40-90 (1990).

\noindent
[2] M. Gomes, R.S. Mendes, R.F. Ribeiro and A.J. da Silva, Phys. Rev.
{\bf D43} (1991) 3516.

\noindent
[3] S.J. Hands, Phys. Rev. {\bf D51} (1995) 5816.

\noindent
[4] B. Rosenstein, B.J. Warr and S.H. Park, Phys. Rep. {\bf205} (1991)
59.

\noindent
[5] S.J. Hands, A. Koci\'c and J.B. Kogut, Ann. Phys. {\bf224} (1993)
29.

\noindent
[6] G. Gat, A. Kovner and B. Rosenstein, Nucl. Phys. {\bf B5.5} (1992)
76.

\noindent
[7] R.D. Pisarski, Phys. Rev. {\bf D29} (1984) 2423;\hfill\break
T.W. Appelquist, M. Bowick, D. Karabali and L.C.R. Wijewardhana,
Phys. Rev. {\bf D33} (1986) 3704.

\noindent
[8] D.Espriu, A. Palanques-Mestre, P. Pascual and R. Tarrach, Z. Phys.
{\bf C13} (1982) 153;\hfill\break
A. Palanques-Mestre and P. Pascual, Comm. Math. Phys. {\bf95} (1984)
277.

\noindent
[9] D.K. Hong and S.H. Park, Phys. Rev. {\bf D49} (1994) 5507.

\noindent
[10] T. Itoh, Y. Kim, M. Sugiura and K. Yamawaki, Prog. Theor. Phys.
{\bf93} (1995) 417.

\noindent
[11] P.I. Fomin, V.P. Gusynin, V.A. Miranskii and Yu.A. Sitenko, Riv.
Nuovo Cimento {\bf6} (1983) 1;\hfill\break
V.A. Miranskii, Nuovo Cimento {\bf90A} (1985) 149.

\noindent
[12] K.-I. Kondo, Nucl. Phys. {\bf B450} (1995) 251.

\noindent
[13] M.R. Pennington and S.P. Webb, Brookhaven preprint BNL-40886
(1988);\hfill\break
D. Atkinson, P.W. Johnson and M.R. Pennington, Brookhaven preprint
BNL-41615 (1988);\hfill\break
M.R.Pennington and D. Walsh, Phys. Lett. {\bf B253} (1991) 246.

\noindent
[14] E. Dagotto, A. Koci\'c and J.B. Kogut, Nucl. Phys. {\bf B334}
(1990) 279;\hfill\break
S.J. Hands and J.B. Kogut, Nucl. Phys. {\bf B5.5} (1990) 455.

\noindent
[15] S.P. Booth, R.D. Kenway and B.J. Pendleton, Phys. Lett. {\bf B228}
(1989) 115.

\noindent
[16] A. Ali Khan, M. G\"ockeler, R. Horsley, P.E.L. Rakow, G. Schierholz
and H. St\"uben, Phys. Rev. {\bf D51} (1995) 5.51.

\noindent
[17] C.J. Burden and A.N. Burkitt, Europhys. Lett. {\bf3} (1987) 545.

\noindent
[18] M. G\"ockeler, R. Horsley, P.E.L. Rakow, G. Schierholz
and R. Sommer, Nucl. Phys. {\bf B371} (1992) 713.

\noindent
[19] E. Dagotto, S.J. Hands, A. Koci\'c, J.B. Kogut, Nucl. Phys. {\bf B347}
(1990) 217.

\noindent
[20] S.J. Hands, A. Koci\'c, J.B. Kogut, R.L. Renken, D.K. Sinclair,
K.C. Wang, Nucl. Phys. {\bf B413} (1994) 503;\hfill\break
A. Koci\'c, J.B. Kogut, K.C. Wang, Nucl. Phys. {\bf B398} (1993) 5.5.

\vfill\eject
\null

\centerline{
\vbox{\tabskip=0pt \offinterlineskip
	\def\tablerule{\noalign{\hrule}}
	\halign to230pt{\vrule#\tabskip=1em plus 2em&
	\hfil#\hfil& \vrule#& \hfil#\hfil& \vrule#&
	\hfil#\hfil& \vrule#\tabskip=0pt\cr\tablerule
	height2pt&\omit&&\omit&&\omit&\cr\tablerule
	height2pt&\omit&&\omit&&\omit&\cr
	&param.&&fit I &&fit II&\cr
	height2pt&\omit&&\omit&&\omit&\cr
	height2pt&\omit&&\omit&&\omit&\cr \tablerule
	height2pt&\omit&&\omit&&\omit&\cr \tablerule
	height2pt&\omit&\omit&\omit&\omit&\omit&\cr
	&\multispan5 \hfil $N_f=2$ \hfil &\cr
	height2pt&\omit&\omit&\omit&\omit&\omit&\cr\tablerule
	height2pt&\omit&&\omit&&\omit&\cr
	&$\beta_c$&&2.03(9)&&1.94(4)&\cr%\tablerule
	&$\delta $&&2.32(23)&&2.68(16)&\cr%\tablerule
	&$\beta  $&&0.71(9)&& -- &\cr%\tablerule
	&$A$&&0.32(5)&&0.37(1)&\cr%\tablerule
	&$B$&&1.91(43)&&2.86(53)&\cr%\tablerule
	&$\chi^2/{\rm d.o.f}$&&2.4&&2.1&\cr
	height2pt&\omit&&\omit&&\omit&\cr\tablerule
	height2pt&\omit&\omit&\omit&\omit&\omit&\cr
	&\multispan5 \hfil $N_f=4$ \hfil &\cr
	height2pt&\omit&\omit&\omit&\omit&\omit&\cr\tablerule
	height2pt&\omit&&\omit&&\omit&\cr
	&$\beta_c$&&0.63(1)&&0.66(1)& \cr%\tablerule
	&$\delta $&&3.67(28)&&3.43(19)& \cr%\tablerule
	&$\beta  $&&0.38(4)&& -- &\cr%\tablerule
	&$A$&&0.78(5)&&0.73(2)&\cr%\tablerule
	&$B$&&7.9(2.8)&&6.4(1.5)&\cr%\tablerule
	&$\chi^2/{\rm d.o.f}$&&3.1&&2.0&\cr
	height2pt&\omit&&\omit&&\omit&\cr\tablerule
	\noalign{\medskip}
	}}
}
\centerline{\bf Table 1}

\centerline{Results from the fits}
\vfill\eject
\null

{\bf Figure Captions}

\noindent
Figure 1:
Diagram contributing to coupling constant renormalisation at
leading order in $1/N_f$.

\noindent
Figure 2: chiral condensate $\sigma$ vs. $1/g^2$ for $N=1,2,3$, corresponding
respectively to $N_f=2,4,6$.

\noindent
Figure 3: chiral condensate $\sigma$ vs. $1/g^2$ for $N_f=2$ and
different values of the bare mass $m$.

\noindent
Figure 4: Fisher plot for $N_f=2$, from data at $\beta=1.6 (\triangle)$,
$1.8 (\triangleleft)$, $2.0 (\nabla)$, $2.2 (\triangleright)$, $2.4 (+)$,
$2.6 (\times)$.

\noindent
\def\bull{\vrule height .9ex width .8ex depth -.1ex}
Figure 5: Fisher plot for $N_f=4$, from data at $\beta=0.5 (\circ)$,
$0.6 (\bull)$, $0.7(\diamond)$, $0.8 (\triangle)$, $0.9 (\triangleleft)$,
$1.0 (\nabla)$, $1.1 (\triangleright)$, $1.2 (+)$, $1.3 (\times)$,
$1.4 (\ast)$.

\noindent
Figure 6: chiral condensate $\sigma$ vs. $m$ for different values of
the coupling (here $\beta\equiv1/g^2$ and is not related to the critical
exponent).

\input psfig
\vfill\eject
\baselineskip = 16pt
\vfill
\vbox{
\centerline{
\psfig{figure=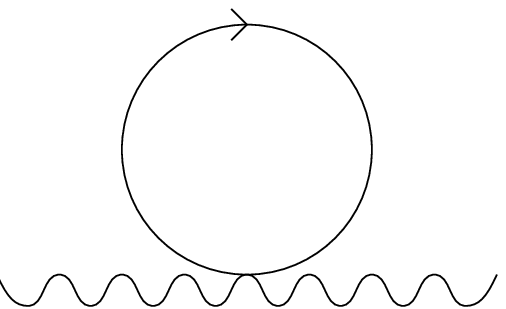,width=4.5in}}
\bigskip\bigskip
\bigskip\bigskip
\bigskip\bigskip
\centerline{\bf Figure 1}
}
\vfill\eject
\vfill
\vbox{
\centerline{
\psfig{figure=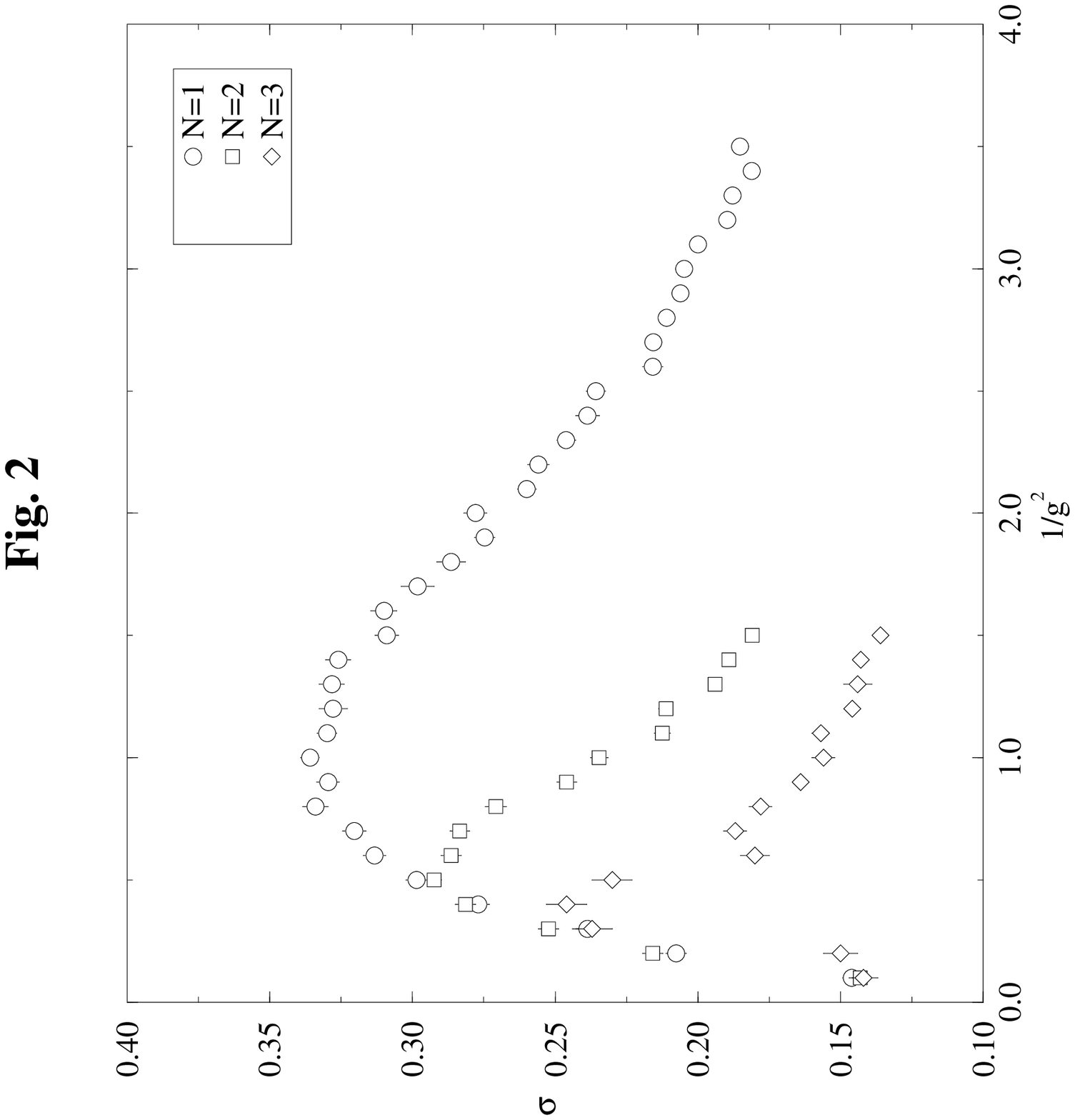,width=5.5in}}
}
\vfill\eject
\vfill
\vbox{
\centerline{
\psfig{figure=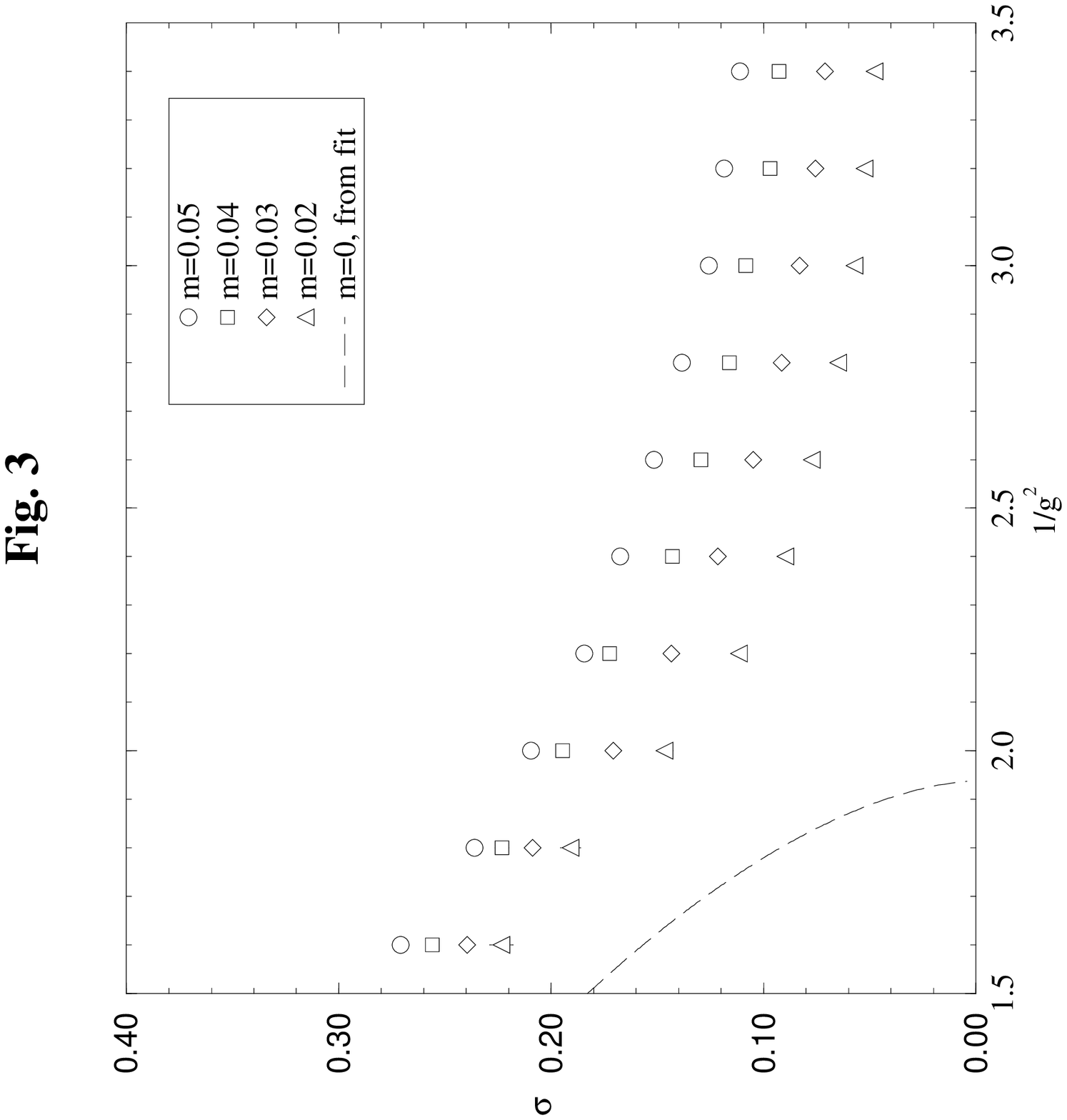,width=5.5in}}
}
\vfill\eject
\vfill
\vbox{
\centerline{
\psfig{figure=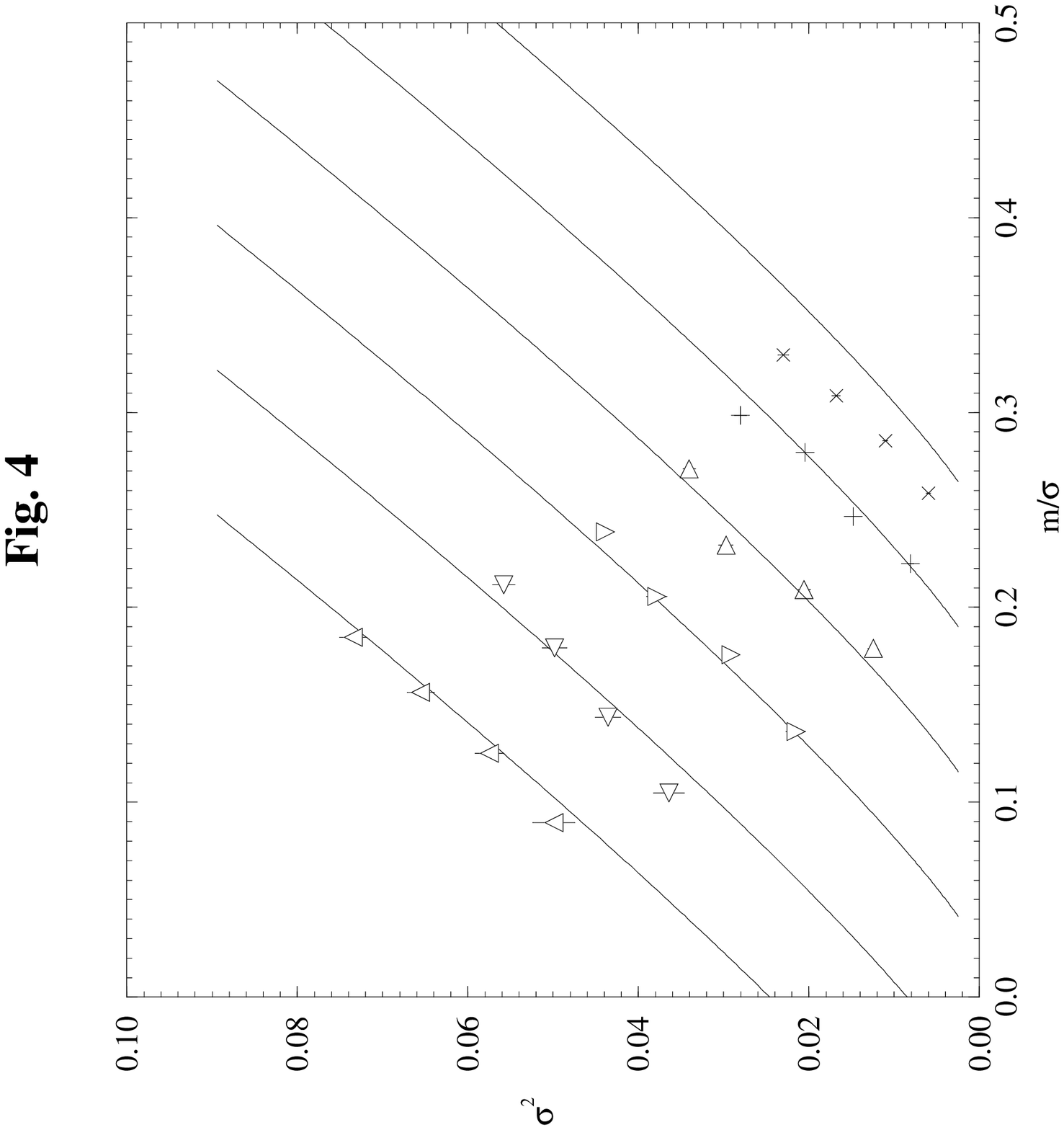,width=5.5in}}
}
\vfill\eject
\vfill
\vbox{
\centerline{
\psfig{figure=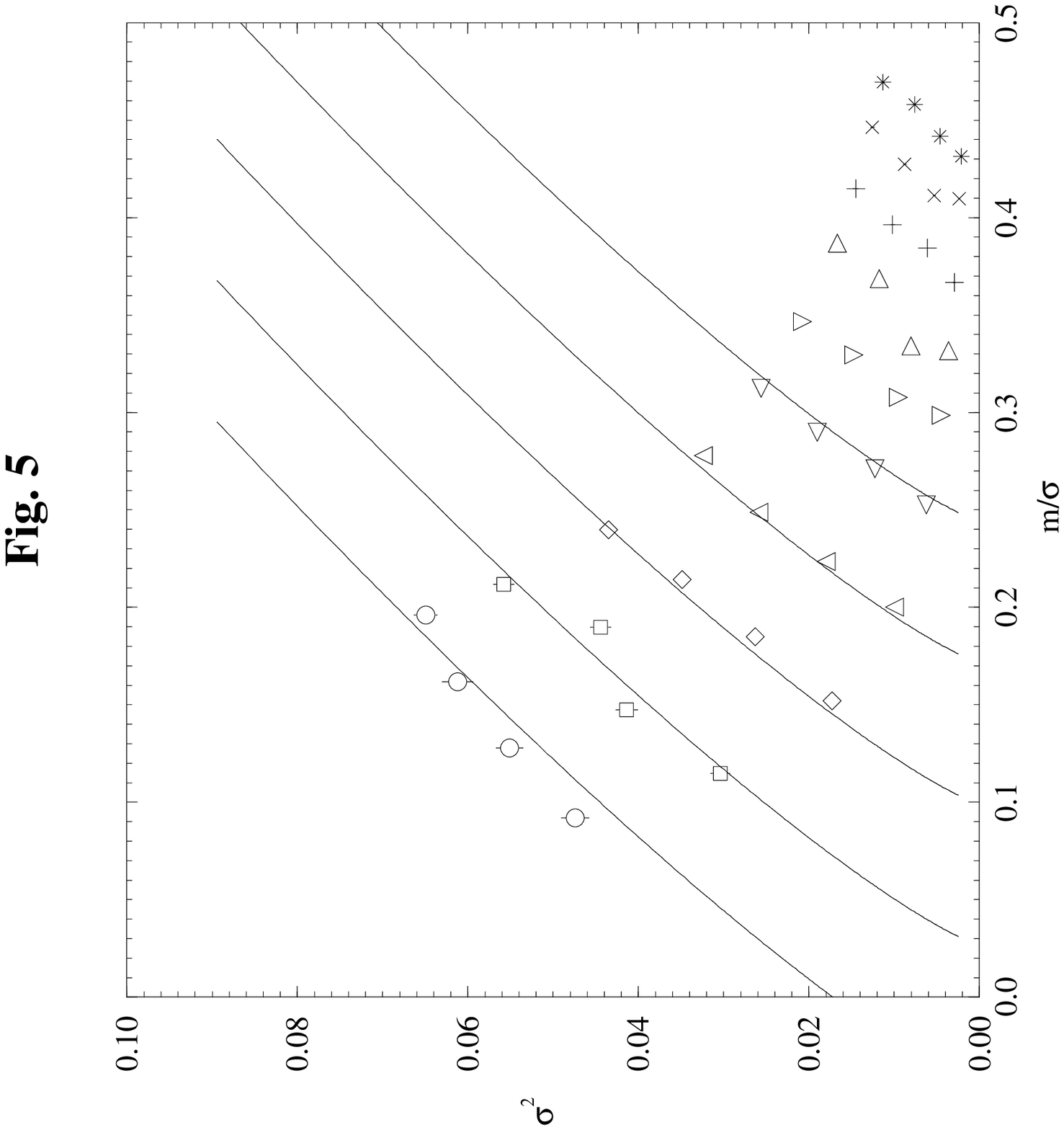,width=5.5in}}
}
\vfill\eject
\vfill
\vbox{
\centerline{
\psfig{figure=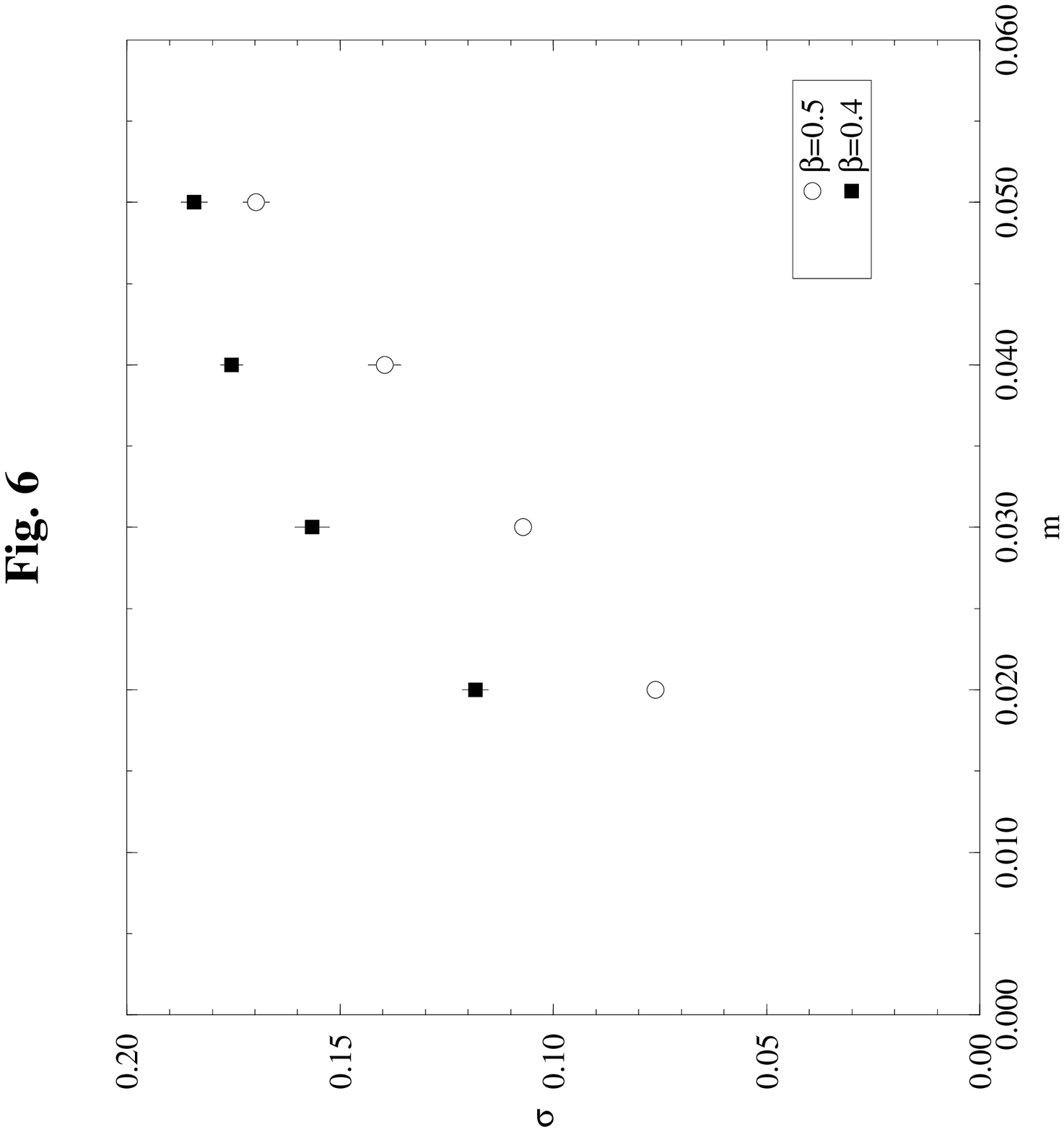,width=5.5in}}
}
\vfill\end